\documentclass[letterpaper,10pt]{article}
\usepackage{osameet3}
\usepackage[font=small]{caption}

\usepackage{amsmath,amssymb}
\usepackage[pdftex,colorlinks=true,bookmarks=false,citecolor=blue,urlcolor=blue]{hyperref} 

\begin{document}
\title{Polarization-Induced Interaction  as a Perturbation within Orbital Angular Momentum (OAM) Modes of an Optical Fiber}
\author{Ramesh Bhandari}
\address{Laboratory for Physical Sciences, 8050 Greenmead Drive, College Park,  Maryland 20740, USA}
\email{rbhandari@lps.umd.edu}
\begin{abstract}
We investigate the role of polarization effects  in orbital angular momentum modes of an optical fiber. Specifically, we revisit the removal of the  degeneracy of the $HE$ and the $EH$ vector modes that exists within the weakly guiding approximation (WGA); within WGA, these two vector modes coalesce into a single spatial (scalar) orbital angular momentum (OAM) mode, defined by a common topological charge  and associated with left or right circular polarization.   While the polarization term of the vector wave equation acting as a Hermitian perturbation operator  accounts for the splitting of the $HE$ and the $EH$ vector modes, its  extension  to explain the accompanying field correction, called the polarization-induced field, encounters hurdles; this field is characterized by a topological charge differing by two in magnitude.   The problems are critically examined with a possible remedy and its implications.
\end{abstract}
\section{Introduction}
 
In the weakly guiding approximation (WGA), the propagation of orbital angular (OAM) modes within an optical  fiber is often described by a scalar wave equation, whose spatial solutions are independent of polarization\cite{snyder, doog, volyar, alex, bhandari}. However, when an externally prepared OAM mode is injected into a fiber, it couples to the vector OAM modes, $HE_{l+1,m}$ or $EH_{l-1,m}$ ($l>1$) according as the polarization of the input mode is left circular (photon spin $S=h/(2\pi)$) or right circular ($S=-h/(2\pi)$). These vector OAM modes  are the solutions of the vector wave equation, expressed in the exponential basis ($e^{\pm il\theta}$) as opposed to the $\cos(l \theta),~\sin(\l\theta)$ basis \cite{snyder}; $h$ is the  Planck's constant; $l$ and $m$ are the topological charge and the radial index, respectively, of the input OAM mode. The vector modes have a slightly different propagation constant from each other as well as the scalar propagation constant. The vector wave equation  includes  a polarization-dependent term, or simply a \emph{polarization} term, which accounts for these slight differences in propagation constants.  The polarization term encompasses two effects, one attributable to the spin-orbit (SO)  interaction  and a second, independent of spin, to a \emph{contact} interaction  in close analogy with   \emph{atomic physics}, where it appears as part of relativistic corrections \cite{davy,schiff, volyar}. While both the SO and the  \emph{contact} interactions contribute to changes in  propagation constant, it is the spin-orbit interaction that causes the breaking of the degeneracy between the $OAM$ modes of the same topological charge but opposite spins (see, e.g.\cite{volyar, alex, raymer}); this degeneracy breaking gives rise  to the nondegenerate $HE_{l+1,m}$ and $EH_{l-1,m}$ modes.
\\\\
In this paper, we critically examine the nature of the polarization effects  as a perturbation  within the OAM  modes of a fiber. In particular, we attempt to understand the presence of the polarization-induced OAM component of the vector modal field as a consequence of the polarization term treated as a perturbation. It is well known that the transverse component of the vector field within a fiber is composed of a modal field, which is the solution of the scalar wave equation (vector wave equation without the polarization term), plus a correction term to account for the absence of the polarization term. The latter field is  polarization-induced  and is of order $\Delta=(n_1^2-n_2^2)/(2n_1^2))$, where $n_1$ and $n_2$ are the refractive indices of the core and the cladding, respectively \cite{snyder}.  More recently,  it was shown that the transverse component of the vector OAM modal field is in general decomposable into two OAM components, one of which is characterized by a topological charge $l$ and the other by a topological charge $l\pm 2$ \cite{bhandari3, ram}. The second component is polarization  induced and of order $\Delta$ (in the \emph{weakly guiding limit}, where $n_1 \rightarrow n_2$, the  parameter $\Delta \rightarrow 0$, and the polarization effects discussed above disappear).  The question then arises if the presence of this polarization-induced component can be explained within the framework of perturbation theory in the same way as the propagation constant splitting between the vector modes and their scalar counterparts.  A recent work \cite{ram2} motivates this connection, but the treatment is incomplete. 
\\\\
In Section 2, we  review the well-known polarization- induced splitting between the $HE_{l+1,m}$ and the $EH_{l-1,m}$ modes as a perturbative effect originating in  the  polarization term of the vector wave equation. Subsequently, in Section 3, we extend this concept to explain the existence of the polarization-induced component, which is of topological charge, $l\pm 2$.  We find that this extension runs into problems of a physical nature,  and suggest a remedy with its ramifications. Section 4 is the summary.
 
\section{Review of polarization effects  as a perturbation}
In what follows, we drop the radial index $m$ for convenience and clarity, and denote the fields of the $HE_{l+1}$ and $EH_{l-1}$ vector modes by $\vec{e}^{(l, 1)}$ and $\vec{e}^{(l, -1)}$, respectively. The corresponding transverse fields, $\vec{e}^{(l,\pm1)}_t$, satisfy the vector wave equation  \cite{snyder}
\begin{equation}
(\nabla^2_t +k^2n^2(r))\vec{e}^{(l, \pm1)}_t+\vec{\nabla_t}(\vec{e}^{(l, \pm1)}_t.\vec{\nabla_t} ln (n^2))=\beta_{l\pm1}^2 \vec{e}^{(l, \pm1)}_t,
\end{equation}
where the third term on the left-hand-side is the \emph{polarization} term; the transverse gradient $\vec{\nabla_t}$ in cylindrical polar coordinates ($r,\theta,z)$ is $ \vec{\nabla_t}=\partial/\partial{r}~\hat{r} +(1/r)\partial/\partial{\theta}~\hat{\theta}$ and the transverse Laplacian $\nabla^2_t=\partial^2/\partial{r^2}+(1/r)\partial/\partial{r}+(1/r^2)\partial^2/\partial{\theta^2}$. The circular fiber is translationally invariant and further assumed to be  axisymmetric. The square of the refractive index is given by $n^2(r)=n_1^2(1-2\Delta f(r))$, where $f(r)$ is the index profile,  $\Delta (=(n_1^2-n_2^2)/(2n_1^2))<<1$ is the profile height parameter, with $n_1$ the refractive index of the core and $n_2$ the refractive index of the cladding as in a step-index fiber  or a  ring-core fiber \cite{sarkar,rusch2}. As an example, $f(r)$ is a step-function for a step-index fiber, being equal to zero for $r\le a$ and equal to 1 for $r>a$, where parameter $a$  is the fiber core radius; for the graded-index fibers, the index profile $f(r)$ is  parabolic, being $r^2$ in the region $r\le a$ and 1 for $r>a$. Parameter $\beta_{l,\pm 1}$ is the propagation constant of the field propagating along the $z$ axis coincident with the fiber axis, and $k=2\pi/\lambda$. 
\\\\
Because $\Delta <<1$, which is usually the case in commercial fibers, the expression $ln(1-2\Delta f(r))\approx -2\Delta f(r)$, which then yields  $\vec{\nabla_t} ln (n^2)=-2\Delta\partial f(r)/\partial r$. Thus, the polarization term in Eq. 1 (in lowest order of $\Delta$)  is proportional to $\Delta$. In  the limit $n_1 \rightarrow n_2$,  $\Delta \rightarrow 0$, implying the polarization effects disappear. Eq. 1 then reduces to the scalar wave equation: 
\begin{equation}
(\nabla^2_t +k^2n^2(r))\vec{\tilde e}^{(l, \pm 1)}_t=\tilde\beta_{l}^2\vec{\tilde e}^{(l, \pm 1)}_t,
\end{equation}
where
\begin{equation}
\vec{\tilde e}^{(l, \pm 1)}_t=O_l({r,\theta})\vec{\epsilon}_{\pm}.
\end{equation}
The polarization-independent spatial amplitude $O_l(r,\theta)$ is given by
\begin{equation}
O_l(r,\theta)=(1/\sqrt{N}_{l})F_{l}(r)e^{ il\theta}.
\end{equation}
  $\vec{\epsilon}_{\pm}=1/\sqrt{2}(\hat{x}\pm i \hat{y})$ are the left/right circularization corresponding respectively to (photon) spin values of $\pm 1$ in units of $h/(2\pi)$. Eq. 3 constitutes an  orbital angular momentum (OAM) modal field with total angular momentum $J (=L+S)$ equal to $(l \pm1)h/(2\pi)$ per photon according as the polarization is left circular (spin $S= +1$)  or right circular (spin $S= -1$); $\tilde\beta_{l} $ is the corresponding scalar propagation constant, and $N_l$ is a normalization constant (the reduction of $\vec{e}^{(l, \pm1)}_t$  to the expression, Eq. 3 is explicitly shown  for the step-index fiber in   \cite{bhandari3}). 
%
\\\\
 Because the polarization term is of order $\Delta$ in Eq. 1, we expect the polarization correction, $\delta\beta_{l,\pm 1}^2=\beta_{l,\pm 1}^2-\tilde{\beta}_l^2$ and the field correction, $ \delta \vec{e}^{(l, \pm1)}_t = \vec{e}^{(l, \pm1)}_t -\vec{\tilde e}^{(l,\pm 1 )}_t$ to be each of order $\Delta$ in the first order of perturbation. Therefore, we  replace $\vec{e_t}^{(l,\pm 1)}$ in Eq. 1 with $\vec{\tilde{e}}_t^{(l,\pm 1)}$ (the unperturbed solution) and substitute Eq. 2 in Eq. 1, we obtain  $(\beta_{l,\pm 1}^2-\tilde{\beta}_l^2)\vec{\tilde e}^{(l,\pm 1 )}_t=\vec{\nabla_t}(\vec{\tilde e}^{(l,\pm 1 )}_t.\vec{\nabla_t} ln (n^2))$.  Taking the scalar product of the above with $\vec{\tilde e}^{(l,\pm 1 )}_t$,  the polarization correction is
\begin{equation}
 \delta\beta_{l,\pm 1}^2=\beta_{l,\pm 1}^2-\tilde{\beta}_l^2=\int_0^{\infty}\int_0^{2\pi}\vec{\tilde{e}}^{*(l,\pm 1)}_t.\vec{\nabla_t}(\vec{\tilde e}^{(l,\pm 1 )}_t.\vec{\nabla_t} ln (n^2)) rdr d\theta,
\end{equation}
where $\vec{\tilde{e}}^{*(l,\pm 1)}_t$ is the complex conjugate of $\vec{\tilde e}^{(l,\pm 1 )}_t$. 
We now substitute  Eqs. 3, 4,  and the expression  
$\vec{\epsilon}_{\pm}=1/\sqrt{2}(\hat{r}\pm i\hat{\theta})e^{\pm i\theta}$ (in polar coordinates) into Eq. 5. Noting further that $n^2=n^2(r)$ and fields  vanish at the end points of the integral, $r=0$ and $r=\infty$,  
we find after  performing integration by parts that
\begin{equation}
\begin{split}
\delta\beta^2_{l,\pm1}&=\frac{\Delta}{N_{l}}\int_0^\infty\int_0^{2\pi}\frac{\partial f(r)}{\partial r}(F_{l}(r)\frac{dF_{l}(r)}{dr}\mp \frac{l}{r}F_{l}^2(r))r drd\theta\\ &=\frac{2\pi\Delta}{N_{l}}\int_0^\infty\frac{\partial f(r)}{\partial r}(F_{l}(r)\frac{dF_{l}(r)}{dr}\mp \frac{l}{r}F_{l}^2(r))r dr, 
\end{split}
\end{equation}
from which $\delta\beta_{(l,\pm1)} \approx \delta\beta^2_{(l,\pm1)}/(2kn_1)$ can be computed. $N_{l}=2\pi\int_0^\infty F^2_{l}(r)r dr$ is the normalization constant (see also \cite {snyder} in the context of \emph{Linearly-Polarized (LP)} modes).   $\delta\beta^2_{(l,\pm1)}$is of order $\Delta$, as expected. The first term in Eq. 6 is identified with the contribution from the \emph{contact} interaction (this designation arises from atomic physics, where the reduction of the Dirac Hamiltonian for the electron in a central field  in the nonrelativisitc case yields, in addition to the spin-orbit interaction term,  an unknown nonclassical term, later identified with the $contact$ interaction between an electron and the atomic nucleus \cite{davy, schiff}).  This contribution is independent of the spin alignment with the OAM, but can be very large, as demonstrated in the case of a step-index fiber \cite{bhandari2}. The second term in Eq. 6 pertains to the spin-orbit interaction, which has a contribution equal in magnitude but opposite in sign for the spin-orbit alignment $(l,1)$  and nonalignment $(l,-1)$ cases. The  same result is reproduced for the  degenerate  states, characterized by $l$ and $S$ values of opposite signs, i.e.,  $\delta\beta^2_{-l,\mp1}= \delta\beta^2_{l,\pm1}$. 
\subsection{Polarization term  as a perturbation operator}
The result of Eq. 5 can be recast  as a matrix element
\begin{equation}
\delta\beta^2_{l, \pm 1}= <\vec{\tilde e}^{(l,\pm 1 )}_t|V|\vec{\tilde e}^{(l,\pm 1 )}_t>,
\end{equation}
where the  operator $V$ represents the perturbative effects of the \emph{polarization} term in Eq. 1. 
%
Subsequently,  entire Eq. 1 in conjunction with Eqs. 2 and 7 may be expressed as 
\begin{equation}
<\vec{\tilde e}^{(l,\pm 1 )}_t|(H_0+V)|\vec{\tilde e}^{(l,\pm 1 )}_t>=\tilde\beta_{l}^2+\delta\beta^2_{(l,\pm1)},
\end{equation}
where $H_0=\nabla^2_t +k^2n^2(r)$; the same result holds for the degenerate OAM-spin combinations: $(-l, -1)$ and $(-l,+1)$.  The operator  $H_0+V$ is Hermitian (as expected) in the space defined by the four OAM states with the OAM-spin combinations: $(l, 1), (-l,1), (-l,-1), (-l,1)$. The  off diagonal elements of the  $4~x~4$ matrix in this space  are each zero; for example,  $ \delta\beta^2_{(-l,1)(l, 1)}=<\vec{\tilde e}^{(-l, 1 )}_t|V|\vec{\tilde e}^{(l, 1 )}_t>$ given by $\int_0^{\infty}\int_0^{2\pi}\vec{\tilde{e}}^{*(-l,1)}_t.\vec{\nabla_t}(\vec{\tilde e}^{(l, 1 )}_t.\vec{\nabla_t} ln (n^2)) rdr d\theta$, following Eq. 5,  evaluates to zero, and so do others. The $4~x~4~H_0+V$ matrix is therefore diagonal; each diagonal element corresponds to a different value of the total angular momentum, $J=L+S$. 
\section{Extension of Perturbation Theory to Include the Polarization-Induced Fields}
We noted above that the polarization effects also induce a field correction, $ \delta \vec{e}^{(l, \pm1)}_t$. This field correction is a state of topological charge, $l\pm 2$, associated with  polarization, $\vec{\epsilon}_{\mp}$.  This is easily seen from a generic form of the  transverse vector modal field of Eq. 1 :  $\vec{e}^{(l, \pm1)}_t=(e^{(l, \pm 1 )}_r(r)\hat{r}+e^{(l, \pm 1)}_{\theta}(r)\hat{\theta})e^{i(l\pm 1)\theta}$, which, upon substitution of $\hat{r}= \hat{x}\cos\theta +\hat{y}\sin\theta$ and $\hat{\theta}=-\hat{x}\sin\theta+\hat{y}\cos\theta$,  becomes 
\begin{equation}
\vec{e}^{(l, \pm1)}_t=(1/{\sqrt{2})}\Big[\vec{\epsilon}_{\pm} e^{il\theta}\big(e^{(l+ \pm 1)}_r(r)\mp i e^{(l, \pm 1)}_\theta(r)\big) + \vec{\epsilon}_{\mp} e^{i(l \pm2)\theta}\big(e^{(l, \pm 1)}_r(r)\pm i e^{(l,\pm 1)}_\theta(r)\big)\Big].
\end{equation}
The field has two components. The second component (of opposite polarization, $\vec{\epsilon}_{\mp}$)  is polarization induced and of order $\Delta$, as discussed in Section 2. Therefore, in the limit $n_1\rightarrow n_2$, this component goes to zero while the  first component reduces to the traditional scalar mode (see Eq. 3).  
%
A detailed analysis of Eq. 9 for a multilayered fiber \cite{bhandari3} reveals the novel result that the second term corresponds to a \emph{modified} scalar mode of topological charge $l\pm 2$, but for now we will work within the framework of perturbation theory based on the traditional  OAM modes of Eqs. 3 and 4 (solutions of the scalar wave equation, Eq. 2).  Function $F_l(r)$ in Eq. 4  is the familiar Bessel function for a step-index fiber and a combination of appropriate Bessel functions in the case of the ring-core fiber \cite{sarkar, rusch2, bhandari4}. To extend the above perturbative model into the space incorporating the states $(l \pm 2, \vec{\epsilon}_{\mp})$, we must consider matrix elements such as $<\vec{\tilde e}^{(l+2, -1 )}_t|H_0+V|\vec{\tilde e}^{(l, +1 )}_t>$ and $<\vec{\tilde e}^{(l, 1 )}_t|H_0+V|\vec{\tilde e}^{(l+2,-1 )}_t>$ for the $J=L+S=l+1$ case, and similarly, $<\vec{\tilde e}^{(l-2, 1 )}_t|H_0+V|\vec{\tilde e}^{(l, -1 )}_t>$ and $<\vec{\tilde e}^{(l, -1 )}_t|H_0+V|\vec{\tilde e}^{(l-2,1 )}_t>$ for the $J=L+S=l-1$ case;  integrations over $r$ and $\theta$ are performed as in Eq. 5 to evaluate them.
\\\\
In the above expanded space and in particular its subspace spanned by the two OAM states: $(l,1)$ and $(l+2, -1)$, each corresponding to total angular momentum, $J=l+1$, we now consider  the four matrix elements: $<\vec{\tilde e}^{(l,1 )}_t|H_0+V|\vec{\tilde e}^{(l, 1 )}_t>,~ <\vec{\tilde e}^{(l+2, -1 )}_t|H_0+V|\vec{\tilde e}^{(l+2, -1 )}_t>,~<\vec{\tilde e}^{(l+2, -1 )}_t|H_0+V|\vec{\tilde e}^{(l, 1 )}_t>$, and $<\vec{\tilde e}^{(l, 1 )}_t|H_0+V|\vec{\tilde e}^{(l+2, -1 )}_t>$ . The matrix elements respectively equal  $\tilde{\beta}_l^2+\delta\beta^2_{l},~\tilde{\beta}_{l+2}^2+\delta\beta^2_{l+2},~\delta\beta^2_{l+2,l}, ~\delta\beta^2_{l,l+2}$ (in this abbreviated notation, where  we suppress spin $S$, index $l$  is equivalent to $(l, 1)$, and index $l+2$ is equivalent to $ (l+2,-1)$). The  operator $H = H_0+V$ is then represented by the  $2~x~2$ matrix
\begin{equation}
H=
\begin{bmatrix}
\tilde{\beta}_l^2+\delta\beta^2_{l}&\delta\beta^2_{l,l+2}\\
\delta\beta^2_{l+2,l}& \tilde{\beta}_{l+2}^2+\delta\beta^2_{l+2}
\end{bmatrix}.
\end{equation}
The above matrix is non-Hermitian because
\begin{equation} 
\begin{split}
\delta\beta^2_{l+2,l}=<\vec{\tilde e}^{(l+2, -1 )}_t|V|\vec{\tilde e}^{(l, 1 )}_t> &= \int_0^{\infty}\int_0^{2\pi}\vec{\tilde{e}}^{*(l+2,-1)}_t.\vec{\nabla_t}(\vec{\tilde e}^{(l, 1 )}_t.\vec{\nabla_t} ln (n^2)) rdr d\theta\\
&=\frac{2\pi\Delta}{\sqrt{N_{l}N_{l+2}}}\int_0^\infty\frac{\partial f}{\partial r}F_{l}\Big(\frac{dF_{l+2}}{dr}+ \frac{l+2}{r}F_{l+2}\Big)r dr
\end{split}
\end{equation}
%
is \emph{not} equal to 
\begin{equation} 
\begin{split}
\delta\beta^2_{l,l+2}=<\vec{\tilde e}^{(l, 1 )}_t|V|\vec{\tilde e}^{(l+2, -1 )}_t> &= \int_0^{\infty}\int_0^{2\pi}\vec{\tilde{e}}^{*(l,1)}_t.\vec{\nabla_t}(\vec{\tilde e}^{(l+2, -1 )}_t.\vec{\nabla_t} ln (n^2)) rdr d\theta\\
&=\frac{2\pi\Delta}{\sqrt{N_{l}N_{l+2}}}\int_0^\infty\frac{\partial f}{\partial r}F_{l+2}\Big(\frac{dF_{l}}{dr}-\frac{l}{r}F_{l}\Big)r dr~;
\end{split}
\end{equation}
the integrations here follow the same steps as in Eq. 5. Same conclusion is reached for $J=l-1$ in a subspace spanned by the two OAM states; $(l,-1)$ and $(l-2, 1)$. This feature of the $H$ matrix is highly  undesirable as it leads to the violation of the orthogonality of the eigenstates. In the next section, we explicitly demonstrate the nonorthogonality of the eigenstates of $H$ and point out the related unphysical implications.
\subsection{Consequences of the non-Hermitian matrix H}
We examine the  orthogonality of the eigenstates  of matrix $H$. Its diagonalization yields the eigenvalues: $\lambda_1\approx \beta^{'2}_{l}+\kappa^2/\beta'_D,~\lambda_2\approx\beta^{'2}_{l+2}-\kappa^2/\beta'_D$, where $\kappa^2=\delta\beta_{l,l+2}^{2}\delta\beta_{l+2,l}^{2}, ~\beta_l^{'2}=\tilde{\beta}_l^2+\delta\beta_l^2,~\beta_{l+2}^{'2}=\tilde{\beta}_{l+2}^2+\delta\beta_{l+2}^2; \beta'_D= \tilde{\beta}_l^{'2}-\tilde{\beta}_{l+2}^{'2}\approx \tilde{\beta}_l^2-\tilde{\beta}_{l+2}^2$. The eigenstate corresponding to  eigenvalue $\lambda_1$ is   a $2~x~1$ column vector: $\psi_1=[1~~\delta\beta_{l,l+2}^{2}/\beta'_D]^T$, representing the field, $\vec{e}_t^{(l, +1)}$. In terms of the field notation,  
\begin{equation}
\vec{e}_t^{(l, +1)}= O_l\epsilon_+ +\Big(\frac{\delta\beta_{l,l+2}^{2}}{\tilde{\beta}^2_l-\tilde{\beta}^2_{l+2}}\Big)O_{l+2}\epsilon_-.
\end{equation}
%
Similarly, the eigenstate corresponding to the eigenvalue $\lambda_2$  is found to be $\psi_2=[-\delta\beta_{l+2,l}^{2}/\beta'_D~~ 1]^T$, which translates to
\begin{equation}
\vec{e}_t^{(l+2, -1)}= O_{l+2}\epsilon_- -\Big(\frac{\delta\beta_{l+2,l}^{2}}{\tilde{\beta}^2_l-\tilde{\beta}^2_{l+2}}\Big)O_{l}\epsilon_+.
\end{equation}
 The scalar product of the fields in Eqs. 13 and 14
\begin{equation}
<\vec{e}_t^{(l+2, -1)}|\vec{e}^{(l, -1)}>=\frac{\delta\beta_{l,l+2}^{2}-\delta\beta_{l+2,l}^{2}}{\tilde{\beta}^2_l-\tilde{\beta}_{l+2}^2}\ne 0
\end{equation}
because $\delta\beta_{l+2,l}^{2}\ne\delta\beta_{l,l+2}^{2}$ as  seen in Section 3.
\\\\
The above result  violates the condition of orthogonality of the modes in a fiber. Physically, this implies that  an $ (l,1)$ mode and an $(l+2, -1)$ mode  injected into an ideal, straight fiber (with no imperfections) will not travel as  independent states, but couple into each other.   \emph{This has the undesired and misleading  implication of crosstalk between two independent modes, $HE_{l+1}$ and $EH_{l+1}$ in an optical fiber}. 
\\\\
Similarly, in the $J=l-1$ case, composed of states, $O_l\epsilon_-$ and $O_{l-2}\epsilon_+$, we obtain the results:
\begin{equation}
\vec{e}_t^{(l, -1)}= O_l\epsilon_- +\Big(\frac{\delta\beta_{l,l-2}^{2}}{\tilde{\beta}^2_l-\tilde{\beta}^2_{l-2}}\Big)O_{l-2}\epsilon_+.
\end{equation}
%
%
\begin{equation}
\vec{e}_t^{(l-2, 1)}= O_{l-2}\epsilon_+ -\Big(\frac{\delta\beta_{l-2,l}^{2}}{\tilde{\beta}^2_l-\tilde{\beta}^2_{l-2}}\Big)O_{l}\epsilon_-.
\end{equation}
The scalar product here is also not equal to zero because $\delta\beta_{l,l-2}^{2}\ne \delta\beta_{l-2,l}^{2}$. This leads to the spurious result that the modes $EH_{l-1}$ and $HE_{l-1}$ mix within a fiber.  
Prior works \cite{ram2} are incomplete as they overlook the issue of non-Hermiticity and the associated problem of non-orthogonality and its erroneous implications, as demonstrated here. 
%
\subsection{Remedy for the non-Hermitian matrix $H$}
A way to get around the above difficulty (that is, to preserve orthogonality) might be to rewrite the non-Hermitian matrix $H$ defined in the space spanned  by the  states: $(l, 1), (l+2,-1)$ of the same $J$ value, $l+1$,  as 
\begin{equation}
H=M+R
\end{equation}
where
\begin{equation}
M=(H+H^{\dagger})/2
\end{equation}
is a Hermitian matrix and
\begin{equation}
R=(H-H^{\dagger})/2
\end{equation}
is an anti-Hermitian matrix (with diagonal elements equal to zero).  We subsequently consider the eigenstates of $M$, treating the matrix $R$ as an error matrix. 
\\\\
Matrix $M$  has each of its two  off-diagonal elements equal to $\delta\beta_{l,l+2}^{'2}=(\delta\beta^2_{l,l+2}+\delta\beta^2_{l+2,l})/2$.  The eigenstates of $M$ have the same expression as those in Eqs. 13 and 14, except that $\delta\beta^2_{l,l+2}$ and $\delta\beta^2_{l+2,l}$ are each replaced with their mean value,  $\delta\beta_{l,l+2}^{'2}$. Consequently, the right-hand-side of Eq.15 now equals zero, as expected for the eigenstates of a Hermitian matrix. For the  $J=l-1$ case, where the space is spanned by the states, $(l,-1), (l-2, 1)$, similar results obtain.
\\\\
Hitherto, we had suppressed the radial index $m$. We assume now that  index m equals 1 for the $(l,\pm 1)$ state, while the state $(l\pm 2,  \mp 1)$ is permitted to have  any value $(1,2....p)$  for the radial index denoted $m'$ (the derivation above tacitly  assumed one specific value of $m'$).  Allowing for more than one value of $m'$ leads to the expansion of the $2~x~2$ matrix $H$ into a $(p+1)~x~(p+1)$ matrix, where $p$ denotes the maximum permitted value of $m'$; parameter $p$ is fiber dependent. It can then be shown that the eigenstates specified in Eqs. 13 and 14 and Eqs. 16 and 17 involve a summation over $m'$ on the second term of the right hand side from $m'=1$   to $ m'=p$.  This summation over $m'$ is consistent with a perturbation series expansion over a  complete set of states, labeled by $m'$ for fixed topological charge. Note further, in Eqs. 11 and 12, $F_l$ will correspond to $m=1$ and $F_{l+2}\rightarrow F_{l+2}^{(m')}$ will be $m'$ dependent. 
\subsubsection{Error matrix $R$}
The  matrix $R$ in Eq. 18  may  be deemed an error matrix (Eq. 20) as it represents the deviation from full  Hermiticity  now captured in the Hermitian matrix $M$. More specifically, error E may be defined as the ratio of the magnitude of the  off-diagonal element of the $R$ matrix ($=(\delta\beta^2_{l,l+2}-\delta\beta^2_{l+2,l})/2$) to the magnitude of the off-diagonal element of the $M$ matrix ($=(\delta\beta^2_{l,l+2}+\delta\beta^2_{l+2,l})/2)$).
For the step-index fiber, where $n^2(r)=n_1^2(1-2\Delta f(r))$ and $f(r)$ is a step-function at r=a, where $a$ is the radius of the fiber core (see Section 2), we can easily evaluate the matrix elements, noting that $\partial f(r)/\partial r=\delta(r-a)$,  and $F_l(r)$ is the familiar Bessel function of the first kind, $J_l(p_l r)$; parameter $p_l=\sqrt{k^2n_1^2-\tilde{\beta}_l^2}$, where $\tilde{\beta}_l$  is determined from the characteristic equation \cite{snyder}.  For fiber parameters, $a=25 \mu m, ~n_1=1.461,~ n_2=1.444$, and wavelength $\lambda=1.55 \mu m$, we find the error E   to increase from 10 to 25\% as $l$ is increased from 1 to 4 for $m=1$ and the single case of $m'=1$. We further find that numerically $ \delta\beta^2_{l+2,l}$  in Eq. 10 is of the same sign as $\delta\beta^2_{l,l+2}$, but consistently larger in magnitude, irrespective of the value of $l$ and $m' (\le p)$, implying matrix $H$  is always non-Hermitian. The same holds true in the $J=l-1$ case. 
\subsection{Discussion}
The space spanned by the four states: $\pm l,\pm 1$ (in the $L,S$ notation) has been considered by a number of authors in the past in the context of polarization-induced corrections to the propagation constants of the scalar OAM modes (see, e.g., \cite{volyar, alex, raymer}, and also \cite{snyder} in the context of $ LP$ modes). The reason the perturbation formulation fails here is that venturing outside  this space triggers  a non-Hermitian matrix because the off-diagonal elements like $<\vec{\tilde e}^{(l+2, -1 )}_t|V|\vec{\tilde e}^{(l, 1 )}_t>$  and $<\vec{\tilde e}^{(l, 1 )}_t|V|\vec{\tilde e}^{(l+2, -1 )}_t>$ in the case of $J=l+1$, are not equal (see Eqs. 11 and 12 above). 
Casting the polarization term of Eq. 1 as a polarization interaction operator, $V$, appears to be applicable only to the space  spanned by the above four OAM modes (with a fixed magnitude of topological charge),  where the propagation  constants of the individual modes (of the same topological charge) are changed due to different polarizations ($\vec{\epsilon}_{\pm}$ ), in close analogy to atomic physics where the energy levels are changed on account of $\vec{L}.\vec{S}$ coupling. However, its extension to  generate the polarization-induced field component (see Eqs. 9, 18, and 19) fails as it leads to misleading conclusions and numerical errors that can be significant. In other words,
%
%
 \emph{polarization-induced interaction, strictly speaking,  is a perturbation internal to an OAM mode within a fiber. While it successfully accounts for the modal propagation splitting due to the spin orbit interaction, it cannot  be treated as an external perturbation acting on the (primary)  OAM field of the vector mode to generate the accompanying (secondary) polarization-induced OAM component.} In fact, starting from Eq. 9, it is explicitly shown in \cite{bhandari3} that the  second component in Eq. 9  in fact  corresponds to a \emph{ modified} OAM field, where the intensity pattern of the traditional OAM mode, e.g., the radius of the intensity donut ring, expands or shrinks, depending upon whether the polarization is aligned or antialigned with the OAM. 
\section{Summary}
We have reviewed the inherent polarization-induced interaction  as a perturbation to explain the differences in the propagation constants between the vector OAM modes $HE_{l+1,m}$ and $EH_{l-1,m}$  in an optical fiber. The transverse fields are eigenstates of the total angular momentum operator, $J=l\pm 1$. Similar results hold for the degenerate vector modes, $HE_{-l-1,m}$ and $EH_{-l+1,m}$, which correspond to $J=-l\mp 1$. Mathematically, the propagation constant differences between the vector modes can   be accounted for by treating the polarization term within the vector wave equation  as a perturbation defined in the space of the four OAM states: $(\pm l, \pm 1)$. However,  its extension as a perturbation into the expanded space incorporating the polarization-induced OAM states $(l\pm 2, \mp 1)$ generates a non-Hermitian matrix with serious, misleading and  physically incorrect  implications. We conclude therefore that  polarization interaction acts internally within a given vector mode to generate the change in the propagation constant, but cannot strictly be treated as a perturbation operator to account for the polarization-induced field (the secondary component of the vector OAM modal field). Instead, a  straightforward, direct analysis of the analytic expression for the polarization induced field reveals it to be a modified version of the traditional scalar OAM field \cite{bhandari3}.
%
%

\end{document}